\documentclass[fleqn,usenatbib]{mnras}

\usepackage[T1]{fontenc}
\usepackage{ae,aecompl}

\usepackage{graphicx}	
\usepackage{amsmath}	
\usepackage{amssymb}	
\usepackage{threeparttable}
\usepackage{multirow}
\usepackage{url}
\usepackage{multicol}
\usepackage{xcolor}

\newcommand{\mum}{\ifmmode{\rm \mu m}\else{$\mu$m}\fi}
\newcommand{\Msun}{\ensuremath{{\rm M}_{\odot}}}   
\newcommand{\Msuny}{\ensuremath{{\rm M}_{\odot} \, {\rm yr}^{-1}}}  
\newcommand{\chisq}{\ifmmode{\chi^{2} }\else{$\chi^2$}\fi}
\newcommand{\rchisq}{\ifmmode{\chi^{2} }\else{$\chi^2_\nu$}\fi}

\usepackage{newtxtext,newtxmath}

\title[HST medium-band imaging of M32]{Hubble Space Telescope imaging of the compact elliptical galaxy M32 reveals a dearth of carbon stars}

\author[O. C. Jones et al.]{O.~C.~Jones,$^{1}$\thanks{E-mail: olivia.jones@stfc.ac.uk}
M.~L.~Boyer,$^{2}$
I.~McDonald,$^{3,4}$
M.~Meixner,$^{5,6}$
J.~Th.~van~Loon$^{7}$
\\
$^{1}$UK Astronomy Technology Centre, Royal Observatory, Blackford Hill, Edinburgh, EH9 3HJ, UK\\\(\)
$^{2}$Space Telescope Science Institute, 3700 San Martin Drive, Baltimore, MD 21218, USA\\
$^{3}$Jodrell Bank Centre for Astrophysics, School of Physics and Astronomy, University of Manchester, Oxford Road, Manchester M13 9PL, UK\\
$^{4}$The Open University, Walton Hall, Kents Hill, Milton Keynes, MK7 6AA, UK\\
$^{5}$Stratospheric Observatory for Infrared Astronomy, NASA Ames Research Center, MS 232-11, Moffett Field, 94035 CA, USA\\
$^{6}$Jet Propulsion Laboratory, California Institute of Technology, 4800 Oak Grove Dr., Pasadena, CA 91109, USA\\
$^{7}$Lennard-Jones Laboratories, Keele University, ST5 5BG, UK \\\(\)
}

\date{Accepted XXX. Received YYY; in original form ZZZ}

\pubyear{2022}

\begin{document}
\label{firstpage}
\pagerange{\pageref{firstpage}--\pageref{lastpage}}
\maketitle


\begin{abstract}
We present new {\em Hubble Space Telescope} WFC3/IR medium-band photometry of the compact elliptical galaxy M32, chemically resolving its thermally pulsating asymptotic giant branch stars.
We find 2829 M-type stars and 57 C stars. The carbon stars are likely contaminants from M31. If carbon stars are present in M32 they are so in very low numbers. 
The uncorrected C/M ratio is  0.020 $\pm$ 0.003; this drops to less than 0.007 after taking into account contamination from M31. 
As the mean metallicity of M32 is just below solar, this low ratio of C to M stars is unlikely due to a metallicity ceiling for the formation of carbon stars. Instead, the age of the AGB population is likely to be the primary factor. 
The ratio of AGB to RGB stars in M32 is similar to that of the inner disc of M31 which contain stars that formed 1.5--4 Gyr ago. If the M32 population is at the older end of this age then its lack of C-stars may be consistent with a narrow mass range for carbon star formation predicted by some stellar evolution models. 
Applying our chemical classifications to the dusty variable stars identified with {\em Spitzer}, we find that the x-AGB candidates identified with {\em Spitzer} are predominately M-type stars.
This substantially increases the lower limit to the cumulative dust-production rate in M32 to $>$ 1.20 $\times 10^{-5}$ \Msuny.
\end{abstract}





\section{Introduction}
\label{sec:intro}

\noindent The Andromeda (M31) satellite M32 is the prototype compact elliptical galaxy (cE). This rare class of galaxies has extremely high stellar densities, a small effective radius ($r_{\rm{eff}} \sim 0.1-0.7$ kpc) and a high mass (>10$^9$ M$_{\odot}$). The origins of these objects are unclear: are cEs extreme low-mass members of the  elliptical galaxy-class \citep{Kormendy2009,Martinovic2017} or are we seeing a truncated remnant of a much larger spiral galaxy \citep[e.g.][]{Faber1973,Bekki2001,FerreMateu2018,DSouza2018}?

M32 is the closest example of a cE galaxy \citep[785 kpc;][]{Fiorentino2010b, Monachesi2011}, yet its origin has been hotly debated for many decades \citep[e.g.][]{Faber1973,Bekki2001}. Recently, M32 has been implicated as a major actor in the evolutionary history of M31, with \citet{DSouza2018} proposing that M32 is the remnant core of a system which started interacting with M31 roughly 5 Gyr ago. 

Studies of M32's stellar populations have almost entirely focused on either integrated light spectroscopy of the high surface brightness central regions of the system \citep[e.g.][]{Rose2005}, or resolved individual stars with the {\em Hubble} and {\em Spitzer Space Telescopes}, and ground-based adaptive-optics observations at larger galactic radii ($R \gtrsim 1{\buildrel{\,\prime}\over{.}}5$) due to the extreme crowding towards the core \citep[e.g.][]{Grillmair1996, Monachesi2011, Davidge2014, Jones2015a, Jones2021}. 
These approaches have established that M32 has two main stellar populations: an intermediate-age, metal-rich  ([Fe/H] $\simeq$ +0.1) population thought to be $\sim$2--8 Gyr old \citep{Rose2005,Coelho2009}, and an old (8--10 Gyr) stellar population with slightly sub-solar metallicity  ([Fe/H] = --0.2; \citealt{Monachesi2011, Monachesi2012}).  The presence of the intermediate-age population, which accounts for 10--40\% of the overall stellar mass, indicates that M32 had a significant interstellar medium (ISM) a few Gyr in the past \citep{Davidge2000, Monachesi2012, Jones2021}. 

Asymptotic giant branch (AGB) stars are the primary manufacturers of carbon, nitrogen and heavy $s$-process elements \citep{Gallino1998, Karakas2014}. They produce complex molecules (e.g., PAHs; \citealt{Ziurys2006}) and they create substantial quantities of dust \citep{Hoefner2018} which they eject into the ISM at rates up to $10^{-6}$ M$_{\odot}$ yr$^{-1}$ \citep{Srinivasan2016}.
As AGB stars evolve, nuclear-processed material is mixed into the envelope, increasing the surface C/O ratio \citep{Iben1975, Herwig2005}, transforming an O-rich, M-type AGB star (C/O $<$ 1) into a C-type carbon star (C/O $>$ 1); this directly affects the baryonic enrichment of the ISM and the galaxy's integrated spectrum.

The luminous AGB stars observed in M32 are likely the progeny of an intermediate-age population with lifetimes between 0.2--5 Gyr \citep{Monachesi2011}. It is from this intermediate age population that carbon stars are expected to form \citep[e.g.][]{DellAgli2017}.
Attaining C/O $>$ 1 depends on the initial metallicity (oxygen abundance), dredge-up efficiency, and stellar mass-loss rate \citep{Karakas2002}. At solar metallicity, carbon-star formation is predicted to occur in AGB stars with an initial stellar mass of $\sim$1.5--3 M$_{\odot}$ \citep[e.g.][]{Marigo2008, Ventura2010,Karakas2014,DellAgli2017,Ventura2020}. At lower metallicities, this mass range increases and it becomes easier to form carbon stars in intermediate-age populations.  
Thus the C/M ratio is often used as a proxy for the metallicity of intermediate-age populations. 

In the near-IR, molecular features from CN and C$_2$ are present in the 0.9--1.7 $\mu$m spectral region of carbon stars, while water absorption features at $\sim$1.4 $\mu$m are present in O-rich stars. Using the {\em Hubble Space Telescope (HST)} Wide-Field Camera 3 \citep[WFC3;][]{Kimble2008} IR medium-band filters, which are sensitive to these molecular absorption features, \citet{Boyer2015c, Boyer2019} surveyed the inner M31 disc, successfully classifying the AGB populations as either C- or M-type and determining the C/M ratio, which is an established indicator of metallicity \citep[e.g.][]{Cioni2003}. This technique has also been used by \citet{Boyer2017} to characterise the atmospheric chemistry of dust-producing AGB stars in metal-poor dwarf irregular galaxies.  

Using {\em Spitzer} \cite{Jones2015a} identified 110 extreme AGB stars (which represent a small subset of the general AGB population) and over 1500 candidate evolved stars within M32. The ratio of carbon to oxygen-rich AGB stars is unknown, as the chemistry of the
AGB stars are not discernible from their {\em Spitzer} colours.
%
In this paper, we present our newly obtained {\em HST} observations of M32 using the medium-band WFC3/IR filters to unambiguously identify carbon- and oxygen-rich AGB stars and detect the intrinsically faint, early-type AGB stars at the tip of the red giant branch
(TRGB).
Section~\ref{sec:observations} describes the observations and data processing.
In Section~\ref{sec:results} we present our results including the identification of individual  C- and M-type AGB stars in our field. We also discuss contamination by M31 stars. 
In Section~\ref{sec:discussion} the AGB population, the metallicity and other properties are discussed and compared with the existing literature. 
Finally, in Section~\ref{sec:conclusion} we summarise the main results and our conclusions.

\section{Observations and Photometry}
\label{sec:observations}

\begin{figure}
\centering
\includegraphics[trim=0cm 0cm 0cm 0cm, clip=true,width=\columnwidth]{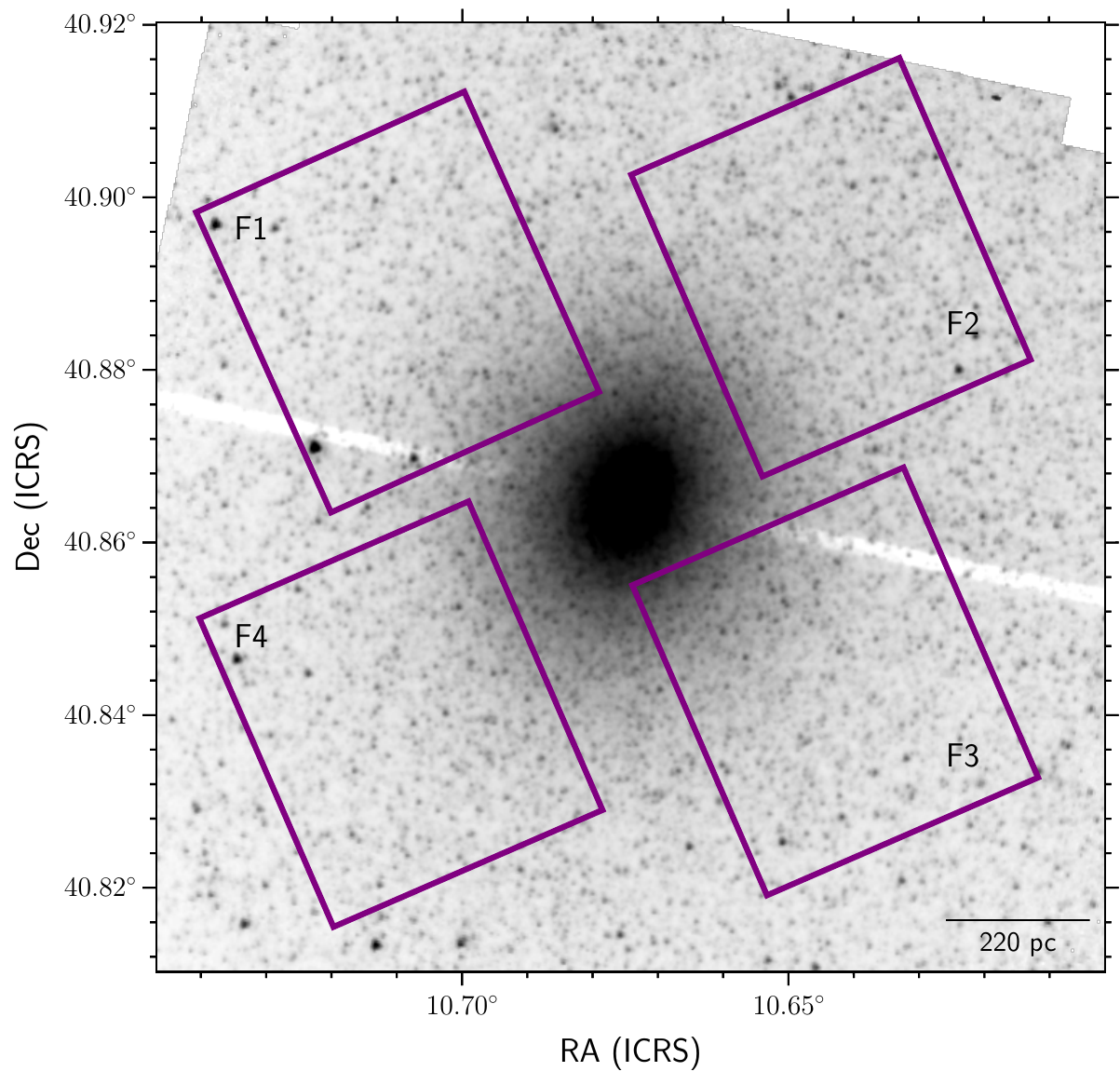}
 \caption{Location of our {\em HST} WFC3/IR footprints towards M32 (solid lines) superimposed on a {\em Spitzer} 3.6$\mu$m mosaic of M32 \citep{Jones2015a} with north pointing up and east to the left. The disc of M31 is towards the top of the image.} 
 \label{fig:M32_obs_cov}
 \end{figure}

We target four fields in M32, with {\em HST's} WFC3/IR medium-band F127M, F139M, and F153M filters in Cycle 27 (GO-15952, PI: Jones). 
Figure~\ref{fig:M32_obs_cov} shows the locations of the $2{\buildrel{\,\prime}\over{.}}1\times 2{\buildrel{\,\prime}\over{.}}3$ WFC3/IR footprints overlaid on a {\em Spitzer} 3.6 $\mu$m image of M32.
These fields represent the best compromise between sampling the bulge of M32 at a point at which crowding should not be a factor versus minimising the M31 background populations.
The WFC3/IR fields cover 60\% of the extreme-AGB stars identified via {\em Spitzer} \citep{Jones2015a}, in addition to two spectroscopically-confirmed carbon stars and 
11 confirmed oxygen-rich stars \citep{Hamren2016}. 

Each field was observed with the F127M, F139M, and F153M filters, at four dither positions using the  {\sc wfc3-ir-dither-box-min} pattern to obtain Nyquist sampling of the point-spread function (PSF)  and minimise image artefacts such as cosmic rays and hot pixels. The total exposure time is 796.9s for F127M and F153M, and 846.9s for F139M at each pointing. 
A summary of the observations is given in Table~\ref{tab:obs_summary}.

\begin{table*}
\centering
\caption{Journal of Observations.}
\label{tab:obs_summary}
\begin{tabular}{@{}ccccccc@{}}
\hline
\hline
Field & RA. & Dec. &  F127M $t_{\rm exp}$ & F139M $t_{\rm exp}$ & F153M $t_{\rm exp}$& Orientation \\
 & (J2000) & (J2000) & (s) & (s) & (s) & (E of N) \\
\hline
F1 & 00h42m49.878s & $+$40d53m12.89s & 796.924 & 846.925 & 796.924 &  $-$65${\buildrel{\circ}\over{.}}$79 \\
F2 & 00h42m33.878s & $+$40d53m27.02s & 796.924 & 846.925 & 796.924 &  $-$66${\buildrel{\circ}\over{.}}$48 \\
F3 & 00h42m33.855s & $+$40d50m33.92s & 796.924 & 846.925 & 796.924 &  $-$66${\buildrel{\circ}\over{.}}$52 \\
F4 & 00h42m49.687s & $+$40d50m20.31s & 796.924 & 846.925 & 796.924 &  $-$66${\buildrel{\circ}\over{.}}$52 \\
\hline
\end{tabular}
\end{table*}

Point sources were extracted from the individual, calibrated, flat-fielded exposures ({\sc flt.fits} files) using the {\sc dolphot 2.0} \citep{Dolphin2000} PSF fitting package. Astrometry for the point source extraction was measured from the drizzled F127M images, and stellar fluxes at these positions in the image stack were obtained using the {\sc dolphot} parameters optimised for IR crowded fields with a wide range of stellar density in M31 by \cite{Williams2014}. 

To ensure the stars in our final catalogue have reliable photometric measurements, we have chosen to retain only point-sources with a signal-to-noise $>5$ in all filters, ${{\rm{\Sigma }}{(\mathrm{Sharp}}_{\lambda })}^{2} < 0.1$, low crowding ${\rm{\Sigma }}\,{(\mathrm{Crowd}}_{\lambda }) < 1.5$ mag, and with a stellar object type. This reduces contamination from extended sources, like background galaxies
and objects whose flux is severely affected by nearby stars. All magnitudes have been transformed into the Vega magnitude system using encircled energy corrections and the photometric zero points from \cite{Kalirai2009a}.
Finally, we correct the photometry for foreground extinction using the dust maps from \cite{Schlafly2011}; we use an $A_V = 0.17$ mag and the \cite{Cardelli1989} extinction curve assuming $R_V = 3.1$. 
%
The final point-source catalogue of 115,295 objects contains only the highest quality photometry; it is available as an online table in the electronic version of this paper and on VizieR. All of the magnitudes we report are in the Vega system, corrected for foreground extinction. The columns in this table are described in Table~\ref{tab:catDescription}. The online catalogue includes all reliable point-sources detected in our FOV including foreground objects.

\begin{table}
\centering
\caption{Numbering, names, and description of the columns present in the master catalogue, which is available as a machine-readable table.}
\label{tab:catDescription}
\begin{tabular}{@{}cll@{}}
\hline
\hline
Column & Name & Description \\
\hline
1 &  ID    &  Unique identification number \\
2 &  RA    &  Right Ascension (J2000)      \\
3 &  DEC   &  Declination (J2000)          \\
4--5 &  F127M, e\_F127M     &  F127M magnitude and 1-$\sigma$ error\\
6--7 &  F139M, e\_F139M     &  F139M magnitude and 1-$\sigma$ error\\
8--9 &  F153M, e\_F153M     &  F153M magnitude and 1-$\sigma$ error\\
10   &  Type                &  AGB classification: C or M \\
\hline
\end{tabular}
\end{table}


\subsection{Artificial star tests}

M32 is extremely compact, and the high surface brightness and stellar crowding poses a major challenge when investigating the resolved stellar content of this elliptical galaxy. 
Our observed fields cover a range in stellar densities from 2 -- 26 sources per arcsec$^2$. 
The rapid fall off in the M32 stellar surface density will cause strong spatial variations in the depth and the accuracy of the photometry, as objects closer to the core are more likely to be affected by crowding.

We performed artificial star tests (ASTs) to evaluate the photometric completeness and uncertainty. This allows us to quantify the effects of blending and crowding on our observed catalogue. Fake stars with magnitudes and colours that fully sample the observed stellar distribution, approximated by the respective PSFs, were randomly injected one at a time into the image stack. These false stars have a limiting magnitude two magnitudes fainter than the faintest detection in the images.  Poisson noise is added to each artificial stars prior to their injection, enabling photometric errors from crowding and background to be constrained independently of the errors due to photon noise.
In total, we injected 70,000 artificial stars and remeasured the photometry as before. A star was considered recovered if it met all the quality cuts described above, and its recovered magnitude was within 0.25 mag  ($|\delta{\rm mag}| < 0.25$) of its input magnitude.

\begin{figure}
\centering
\includegraphics[trim=0cm 0cm 0cm 0cm, clip=true,width=\columnwidth]{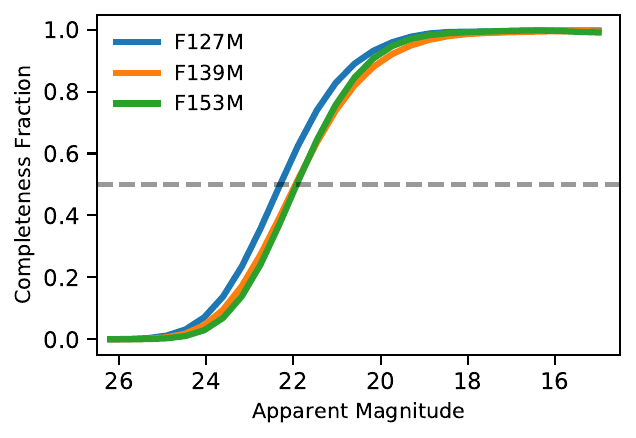}
\includegraphics[trim=0cm 0cm 0cm 0cm, clip=true,width=\columnwidth]{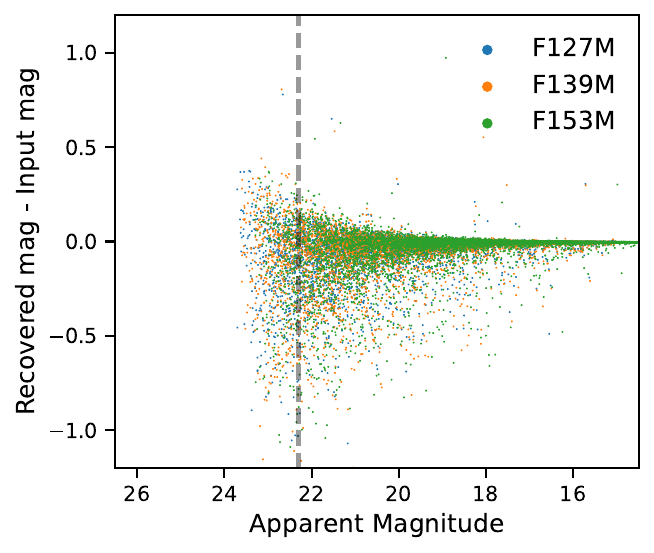}
\includegraphics[trim=0cm 0cm 0cm 0cm, clip=true,width=\columnwidth]{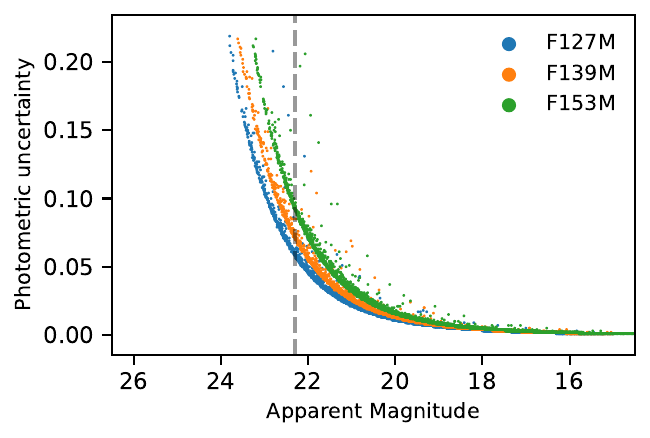}
 \caption{Top: The completeness fraction, typically computed in 0.4 mag intervals, as a function of magnitude for each filter.  The dashed grey line indicates the 50\% completeness limit adopted for our analysis.
 Middle:  Distribution of differences between the true and the recovered magnitudes for artificial stars as a function of the apparent magnitude.
 Bottom: Photometric uncertainties as a function of magnitude for the three {\em HST}/WFC3 medium band filters.} 
 \label{fig:M32_completness}
 \end{figure}

Figure~\ref{fig:M32_completness} shows the completeness fractions and measurements of internal photometric errors from the ASTs as a function of apparent magnitude. The 50\% completeness level is located between 22.3 -- 21.9 mag with the bluer filters reaching deeper magnitudes. 
There is a slight radial dependence in the completeness limit. At radii closer to the core ( at around  $R <65^{\prime\prime}$) photometric completeness starts to decrease, and there is an increased likelihood of stellar blends, preventing detection of M32's intrinsically fainter stars.
Recovered magnitudes tend to be fainter than the input magnitudes of the artificial stars; this effect is more severe for the fainter sources.
This is due to increased uncertainties and over-subtraction of the flux contribution from its neighbours in crowded regions. 
By comparing the true magnitude to the measured magnitude we can obtain a realistic estimate of the internal photometric errors. 
For stars brighter than the TRGB (see Sec.~\ref{sec:TRGB}) the completeness fraction is over 95\%; here the median magnitude differences between the measured and input magnitudes of the false stars are negligible.

\begin{figure}
\centering
\includegraphics[trim=0cm 0cm 0cm 0cm, clip=true,width=\columnwidth]{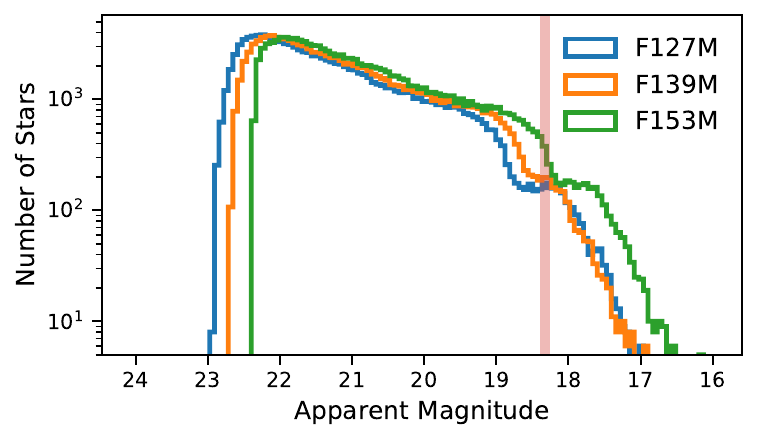}
 \caption{
 Luminosity function of our M32 pointings. The bin-width of 0.06 mag was determined using Knuth's rule for Bayesian bin-width selection \protect\citep{Knuth2006}. The vertical line indicates the measured F153M TRGB position.}
 \label{fig:M32_LumFunction}
 \end{figure}

\subsection{Foreground Milky Way Stars}

To remove foreground Milky Way dwarf stars from our sample, we match our catalogue to the data from {\em Gaia} DR3 \citep{GaiaDR3}. 
At the distance of M32, genuine Galaxy members should either be absent from, or are too faint to have, accurate parallax  measurements \citep{vanderMarel2019, Salomon2021}. 
In total 85 foreground stars matched to the {\em Gaia} DR3 catalogue with $\sigma_\varpi/(\varpi - \varpi_\mathrm{ZP}) < 0.2$ and $(\varpi - \varpi_\mathrm{ZP}) > 0.1$ mas. Where $\varpi$ is the parallax in mas and $\varpi_\mathrm{ZP} = -0.02$  \citep{Lindegren2021}. We do not consider these stars further in our analysis.

We also use the {\sc trilegal} Galaxy model \citep{Girardi2012} to predict the density of foreground stars in our M32 fields. This model predicts $\sim$175 foreground Galactic stars in our 19.32 arcmin$^2$ field. 
Hence, contamination by Galactic foreground stars in our sample is negligible.

\subsection{Mitigating contamination by M31}
\label{sec:contamination}

M32 and M31 are in close proximity, with a projected separation of $24^\prime$ (5.4 kpc; \citealt{Sarajedini2012}). M31 stars are thus a major source of contamination in our M32 sample. We adopt two complementary approaches to statistically correct our M32 C- and M-type star counts to account for this contamination.  

First, we recover the numbers of contaminant C and M-type stars in our M32 field using chemically classified AGB stars in M31 from \citet{Boyer2019}. This survey used the same {\em HST} filters and observing set-up as our programme, to observe 21 fields across M31's disc sampling a wide range in metallicity and age. 
As the ratio of C/M TP-AGB stars decreases rapidly with galactic radius, we fit a linear function to the number of stars as a function of deprojected radius and extrapolate the star counts at M32's deprojected radius of 13.12 kpc. This corresponds to 263 $\pm$ 16 M-type stars and 17 $\pm$ 6 C-stars per WFC3/IR pointing at M32's location. We refer to C and M star counts corrected using these values as `extrapolation corrected'. 
Alternatively, we can consider Field 20 of \citet{Boyer2019} in `Brick 18' of the Panchromatic Hubble Andromeda Treasury (PHAT) program \citep{Dalcanton2012} to be representative of the M31 contamination in our data.  This field has a deprojected distance of 13.68 kpc, is outside of the star-forming arms and contains M31 disc stars; in total, this field has 198 M-type stars and 14 C-stars in a WFC3 pointing. We refer to this M31 correction as `F20 corrected'.

\section{Results}
\label{sec:results}

\subsection{ The Tip of the Red Giant Branch (TRGB)}
\label{sec:TRGB}

Our data should reach below the tip of the red giant branch (TRGB) at the distance of M32 as this value is expected to be comparable to the WFC3 near-IR wide-band TRGB of $\sim$18.5 mag for M31 determined by \cite{Dalcanton2012}. The tip of the first ascent of the red giant branch is a distinct feature in CMDs of old and intermediate-age stellar populations; reached when helium burning is ignited in the electron-degenerate core of low-mass stars.  This helium ignition occurs abruptly at core temperatures of $\sim10^8$ K \citep{Salaris1997}, resulting in a rapid decrease in luminosity as the star transitions to the horizontal branch. 
The subsequent AGB phase of evolution, which reaches higher luminosities than RGB stars is much faster, thus producing a discontinuity in the luminosity function.
As the maximum bolometric luminosity of RGB stars is well constrained \citep{Sweigart1978} the TRGB has been used as a standard candle to precisely measure the distance to nearby galaxies. 

\begin{figure}
\centering
\includegraphics[trim=0cm 0cm 0cm 0cm, clip=true,width=0.8\columnwidth]{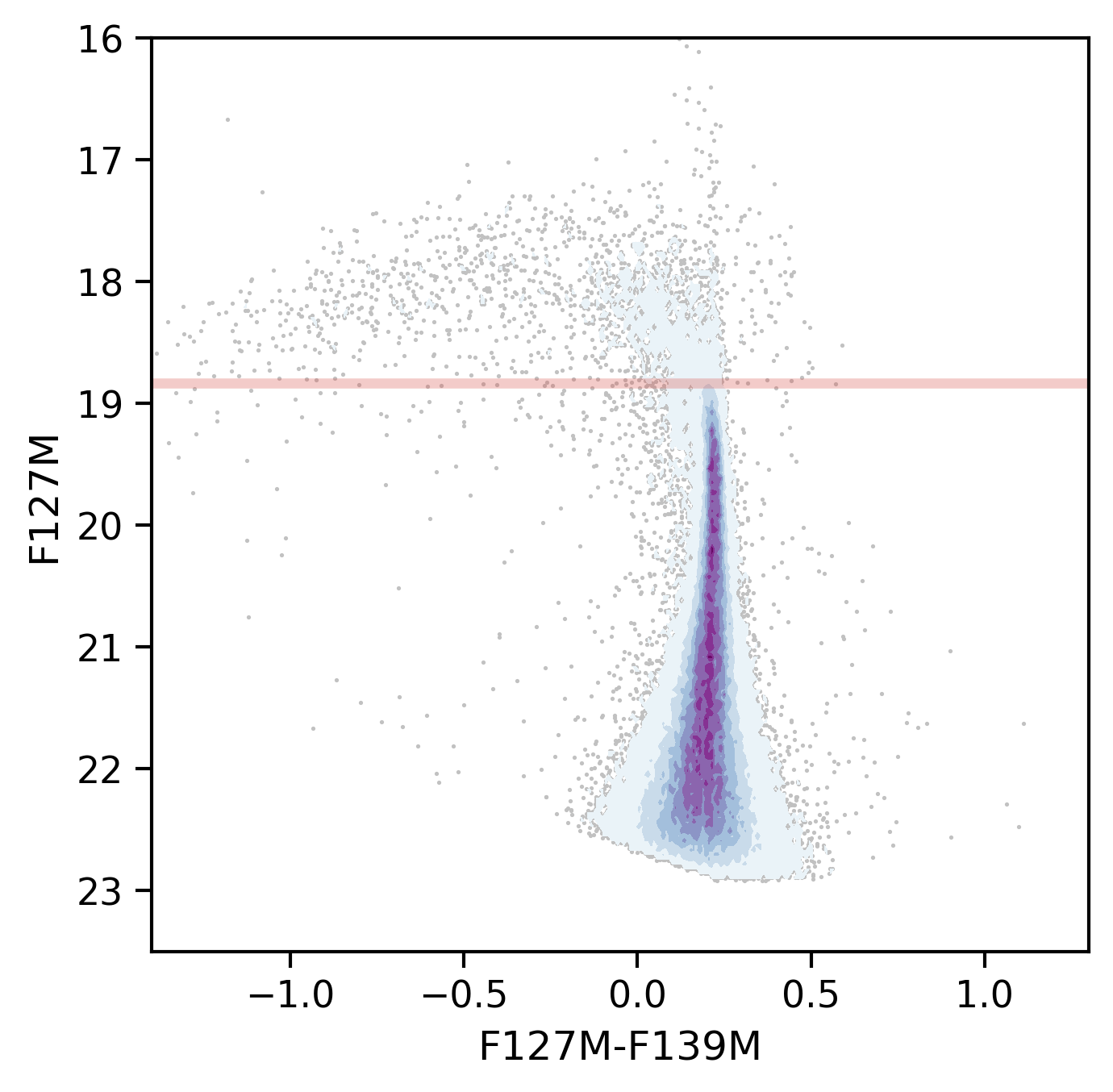}
\includegraphics[trim=0cm 0cm 0cm 0cm, clip=true,width=0.8\columnwidth]{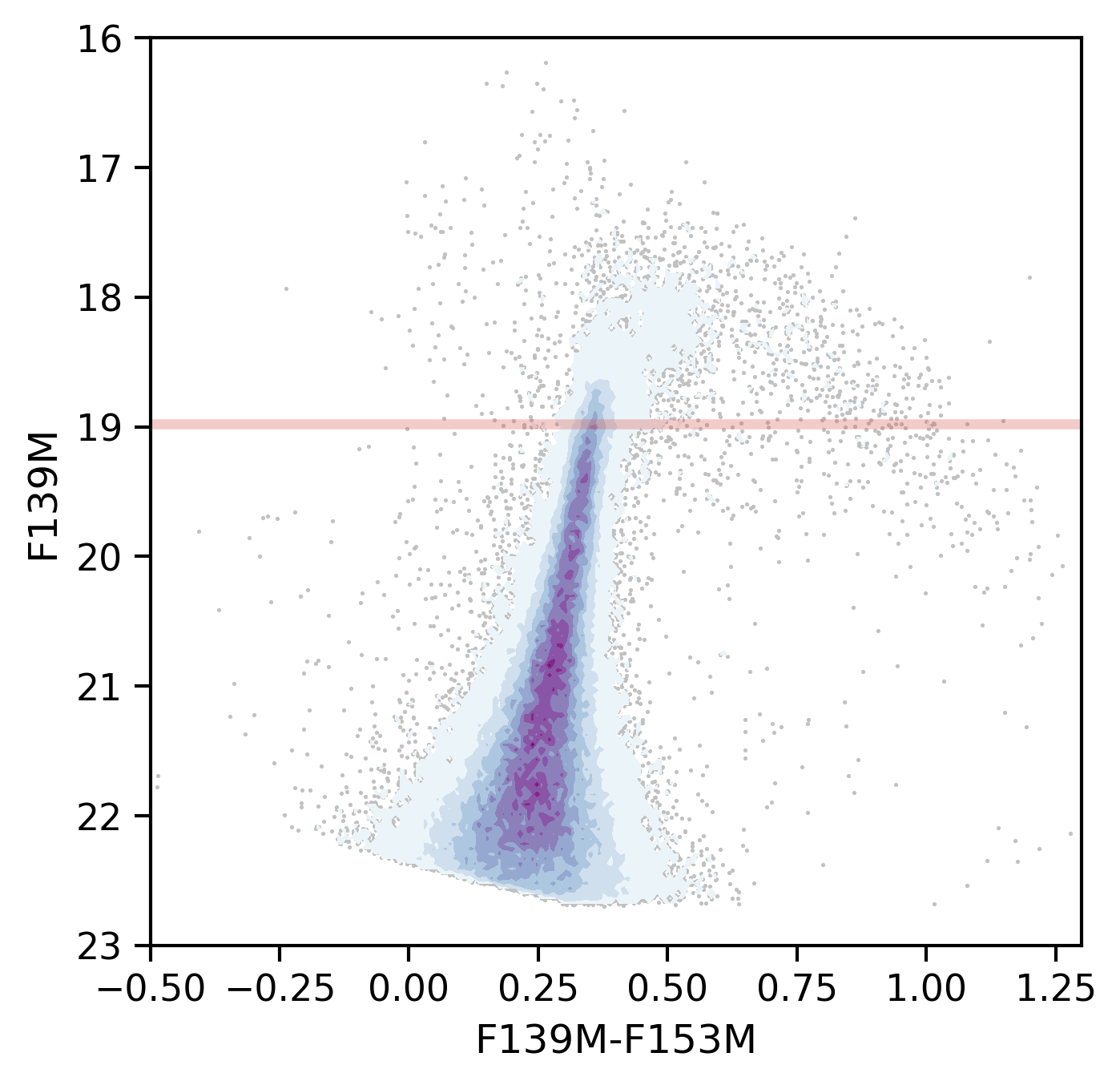}

\includegraphics[trim=0cm 0cm 0cm 0cm, clip=true,width=0.8\columnwidth]{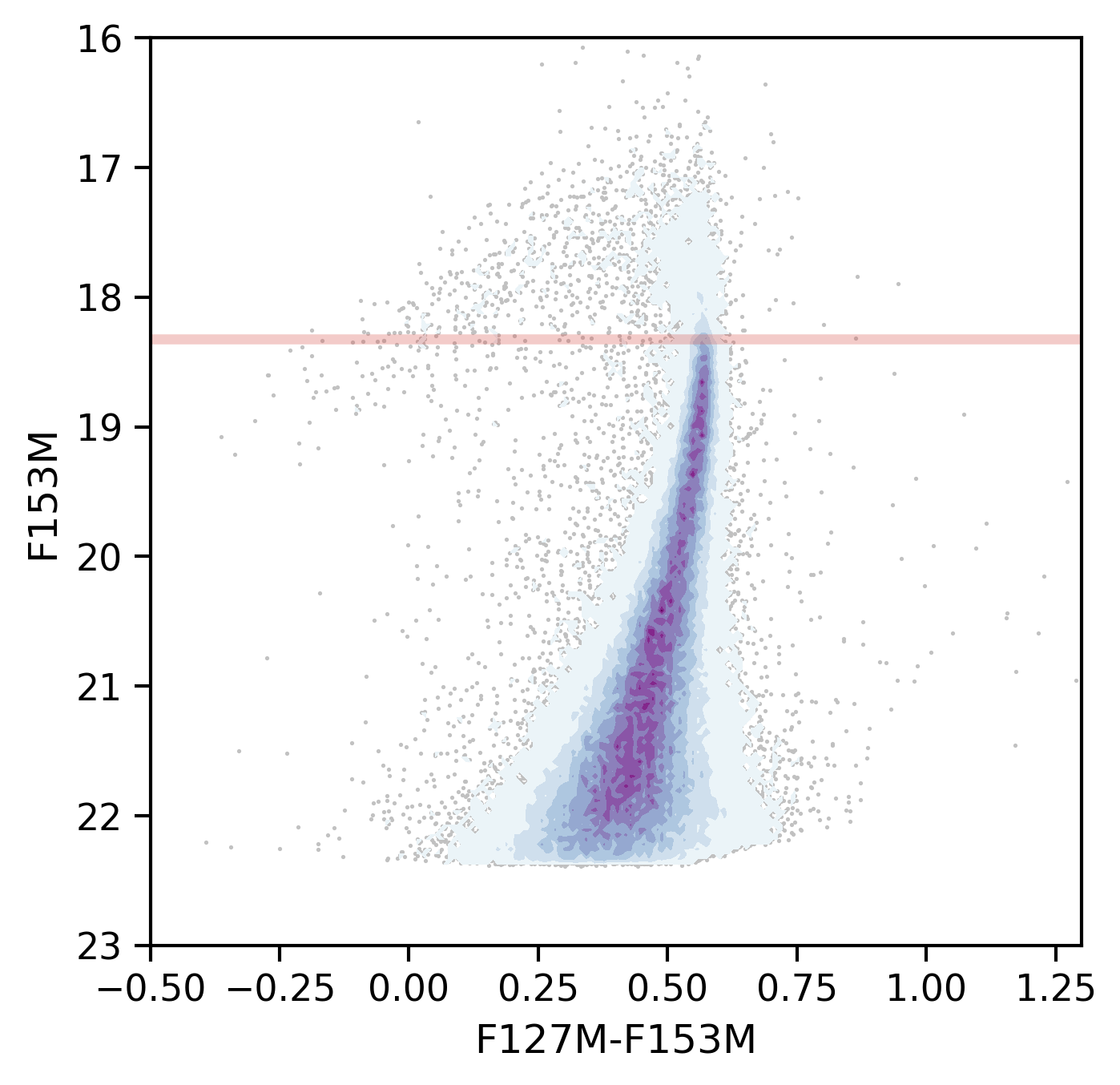}
 \caption{{\em HST} Hess colour--magnitude diagrams, for stars in M32 with ${\rm{S}}/{\rm{N}} > 5$, corrected for foreground extinction. The solid horizontal red line is the location of the tip of the red giant branch (TRGB) in that band identified via the Savitzky-Golay filter. Stars located above the TRGB are considered to be AGB stars. Sources located on the tail of AGB distribution but below the TRGB, due to deep water absorption, are also considered to be AGB stars.}
 \label{fig:M32_CMD}
 \end{figure}

To identify the TRGB in our WFC3/IR photometry, we apply an edge-detection algorithm to measure the discontinuity in our Gaussian-smoothed luminosity functions. The luminosity functions are not corrected for completeness as a function of radius from M32, as this has negligible impact on the bright end of the luminosity function where the TRGB is located. 
For each filter and field, we first select stars with ${\rm{F}}127{\rm{M}}-{\rm{F}}153{\rm{M}} > 0.1$ mag and with photometric errors less than 0.1 mag to construct the luminosity functions (LF) presented in Figure~\ref{fig:M32_LumFunction}. This minimises foreground or main-sequence contaminants and confusion from large photometric uncertainties when determining the TRGB. 
Then, using a Monte Carlo approach that takes into account the observed photometric error distribution by using the minimum and maximum error for a given magnitude bin, we generate 1000 realisations of the Kernel Density Estimation (KDE) used to smooth our data. These errors account for errors in magnitudes due to crowding in addition to errors due to photon noise. Each iteration adds random Gaussian noise based on the photometric uncertainties to the observed luminosity function prior to smoothing with an Epanechnikov kernel. Finally, a Savitzky-Golay filter was applied to each convolved LF to calculate its smoothed first derivative and the magnitude of its minimum, corresponding to the position of the TRGB. The measured TRGBs and their associated error  averaged over 1000 simulations are listed in Table~\ref{tab:TRGB}. We use the TRGB to separate TP-AGB stars, which should be brighter than the TRGB, from our overall sample. 
Figure~\ref{fig:M32_CMD} shows the extinction-corrected CMDs for the combined M32 field of view. These CMDs are populated by a narrow red giant branch (RGB) sequence reaching several magnitudes below the tip of the RGB (TRGB).  As these luminosity functions include all stars in our M32 pointing they contain a significant number of sources from M31 in addition to the M32 population and thus are not suitable for deriving a clean TRGB for the purpose of distance estimates for M32.

\begin{table}
\centering
\caption{Measured apparent magnitudes of the TRGB, using the extinction corrected data.}
\label{tab:TRGB}
\begin{tabular}{@{}cccc@{}}
\hline
\hline
Field & F127M & F139M & F153M \\
 &  (mag) & (mag)  &  (mag)  \\
\hline
F1 &  18.79 $\pm$ 0.15  &   19.02  $\pm$ 0.22 &  18.28  $\pm$ 0.11  \\
F2 &  18.81 $\pm$ 0.12  &   19.00 $\pm$  0.23 &  18.31 $\pm$  0.10   \\
F3 &  18.85 $\pm$ 0.04  &   19.01 $\pm$  0.27 &  18.32 $\pm$  0.06   \\
F4 &  18.74 $\pm$ 0.23  &   18.98 $\pm$  0.22 &  18.26 $\pm$  0.18   \\
\hline
All &   18.83 $\pm$ 0.01  &   18.98 $\pm$  0.26 &  18.32 $\pm$  0.02   \\
\hline
\end{tabular}
\end{table}

\subsection{Identifying carbon- and oxygen-rich AGB Stars}
\label{sec:CM}

We can separate C- and M-type TP-AGB stars using the F127M -- F139M versus F139M -- F153M colour--colour diagram (CCD). This method was first used by \cite{Boyer2013} based on the \cite{Aringer2009,Aringer2016} stellar models, to successfully identify carbon stars in the inner region of M31 with minimal contamination.  
It effectively segregates stars based on the molecular absorption features in their spectra, and is relatively robust against confusion from  both circumstellar and interstellar extinction, as reddening acts along a similar axis to the vector separating the two stellar types. 

To select the C- and M-type AGB stars, we first exclude all point-sources in our catalogue with fluxes below the TRGB in either the F127M or F153M filters. This restricts our sample to thermally-pulsing AGB star candidates. The total number of TP-AGB candidates is given in Table~\ref{tab:AGB_counts}.
In some instances, the presence of deep water absorption features in late M-type stars may depress the stellar flux of TP-AGB stars. To recover these sources we identify stars with blue F127M -- F153M $ <0.17$ mag colours and magnitudes up to 1.5 mag fainter than the TRGB$_{\rm F153M}$. 

\begin{figure}
\centering
\includegraphics[trim=0cm 0cm 0cm 0cm, clip=true,width=\columnwidth]{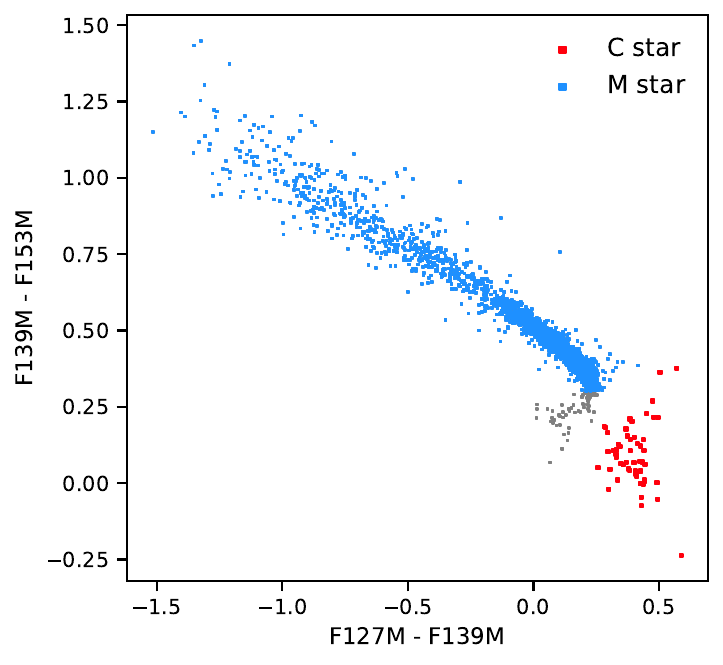}
\includegraphics[trim=0cm 0cm 0cm 0cm, clip=true,width=\columnwidth]{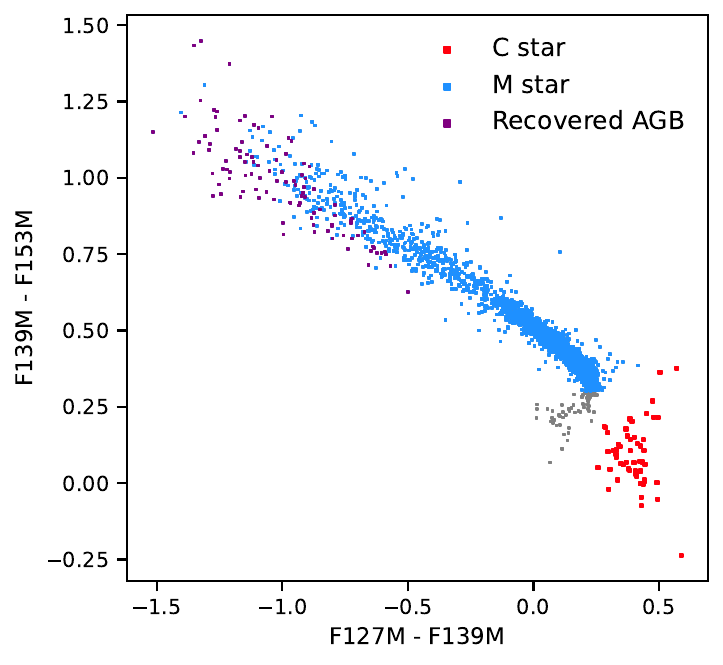}
 \caption{{\em HST} medium-band two-colour diagram to select carbon- and oxygen-rich AGB stars, shown as red and blue squares respectively. The grey points represent K-type stars with T$_{\rm eff} > 3600$ K, or unidentified foreground contaminants. Bottom: As above but highlighting recovered AGB stars (purple) that have a deep water absorption feature and are fainter than the TRGB in all three bands. The number C- and M-type AGB stars identified in this diagram corresponds to the `raw' counts presented in Table~\ref{tab:AGB_counts}.}
 \label{fig:M32_HST_CCD}
 \end{figure}

Figure~\ref{fig:M32_HST_CCD} shows the two-colour diagram used to identify the AGB spectral types.
Based on their position in this colour-colour diagram, we classify the AGB stars as either C- or M-type stars.
Late M-giants can have substantial H$_2$O absorption at 1.3--1.5 \mum\ which is traced by the F139M filter, whilst C-stars have a broad CN$+$C$_2$ feature at $\sim$1.4--1.6\mum\ which falls within the F153M band; thus C- and M-type stars occupy distinct regions in this two-colour diagram. 
The most prominent feature is the branch occupied by M-type stars in the upper-left of the CCD. Here, M-type stars fall along a sequence of increasing water absorption from early M0 stars with the reddest F127M -- F139M colour to late M subtypes with very blue F139M -- F153M colours. The recovered AGB stars with magnitudes fainter than the TRGB (due to deep water absorption features) typically occupy the tail end of this sequence in the top-left corner of the plot. 
Carbon stars occupy the lower-right region of this CCD and are cleanly separated from the other populations. Thus we expect there to be minimal contamination from other sources in our carbon star count, listed in Table~\ref{tab:AGB_counts}. The raw number will include stars from both M32 and M31. After correcting for this contamination no carbon stars belonging to M32 remain (see Section~\ref{sec:Metallicity}).
The hook to the M-star branch seen in the lower-right of the figure is comprised of a combination of K-type, main-sequence, and any remaining foreground stars in our sample. Stars occupying this region of colour space are excluded from our C and M star counts. 

\begin{figure}
\centering
\includegraphics[trim=0cm 0cm 0cm 0cm, clip=true,width=\columnwidth]{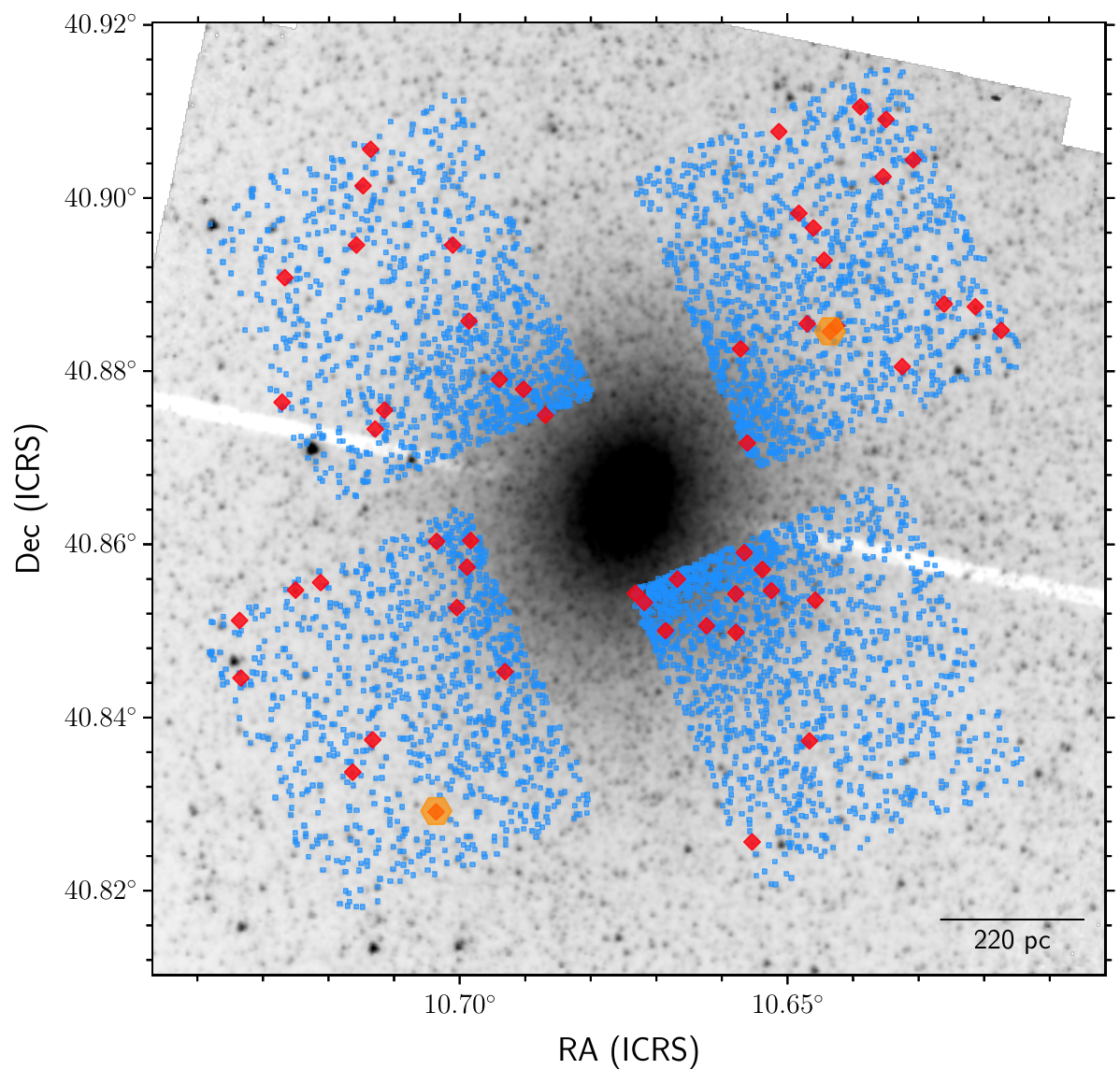}
 \caption{The spatial distribution of sources classified as AGB stars in the M32 {\em HST} FOV before taking into account M31 contamination. Carbon stars are plotted in red, M-stars in blue. The two spectroscopically confirmed carbon stars from \protect\cite{Hamren2016} are plotted as orange hexagons.} 
 \label{fig:M32_Cstar_loc}
 \end{figure}

Optical spectra of 13 stars in our M32 fields are available from the Spectroscopic and Photometric Landscape of Andromeda's Stellar Halo (SPLASH) survey \citep[][]{Guhathakurta2006, Hamren2016}. Within our FOV there are  six M-stars brighter than the TRGB in M32 that \cite{Hamren2016} classified and two carbon stars; we successfully identify all  eight AGB sources as O-rich or C-rich using the {\em HST} two-colour diagram. 
The positions of the HST-identified carbon stars and M-type stars found in M32 are plotted on a {\em Spitzer} map of the galaxy in Figure~\ref{fig:M32_Cstar_loc}. The positions of the two carbon stars identified by \cite{Hamren2016} are also marked.

\begin{table}
\centering
\caption{C- and M-type AGB stars are identified by their {\em HST} near-IR colours. The raw counts are the final number of stars detected. The corrected counts take into account stellar contamination from M31 using two complementary methods. After correcting for contamination no carbon stars belonging to M32 remain (see Section~\ref{sec:Metallicity}). } 
\label{tab:AGB_counts}
\begin{tabular}{@{}lccc@{}}
\hline
\hline
        & Raw  &  M32 Corrected F20 & M32 Corrected   \\
        &      &                    & M31 extrapolation \\
\hline
$N_{\rm TRGB}$  &  3004                 &  \ldots         &  \ldots \\
$N_{\rm M}$     &  2829                 &  2037 $\pm$ 77  &   1777 $\pm$ 64     \\
$N_{\rm C}$     &    57                 &  0 $\pm$ 16     &   $-$12 $\pm$ 24     \\
C/M             &  0.0198 $\pm$  0.0028 &  0.000 $\pm$ 0.008 &   $-$0.007 $\pm$  0.014 \\ 
\hline
\end{tabular}
\end{table}

\section{Discussion}
\label{sec:discussion}

\subsection{C/O ratio and metallicity}
\label{sec:Metallicity}

M32 has a substantial intermediate-age population formed during an extended period of star-forming activity in M32 \citep[e.g.,][and references therein]{Rose1985,Worthey2004a,Davidge2007, Monachesi2011, Monachesi2012, Jones2021}. For these populations, the C/M ratio has long been used as a proxy for its metallicity, as it correlates inversely with the abundance ratio [O/H]. This is due to two main factors: 1) at lower metallicities there are lower natal oxygen abundances and thus fewer carbon atoms need to be dredged up for the transformation from M- to C-type stars; 2) the depth and efficiency of third dredge-up events increase at low metallicity \citep[e.g.,][]{Karakas2002, Karakas2007, Marigo2007, Ventura2020}.

Repeated third dredge-up episodes are limited by the increasing mass-loss rate, which fast becomes the dominant process reducing the mass of the stellar envelope in the later AGB phases. Consequently, a metallicity ceiling is predicted by theoretical models whereby not enough carbon nuclei are transported to the convective envelope to obtain C/O $>$1 before stellar mass loss prevents further dredge-up \citep[e.g.,][]{Marigo2013}; this is expected to occur around [M/H] $>$ $+$0.1, although models for this regime are limited due to a number of complex physical processes which are poorly constrained.  

M32 has very low numbers of carbon stars relative to the size of its TP-AGB population which, prior to contamination subtraction, results in a C/M ratio of 0.020 $\pm$ 0.003. 
However, when using the ratio of C to M stars to determine the true properties of the underlying system, contamination subtraction is crucial, especially in the case of M32 where a large contribution from M31 stars is expected. 
Given this, we have adopted two different approaches for the M31 contamination correction; the first uses the C and M star counts  directly from `Field 20' of \citet{Boyer2019} after adjusting for the difference in area, the second used the \citet{Boyer2019} sample to extrapolate the star counts at M32's deprojected radius (`M31 extrapolation corrected'), these approaches are discussed further in Section~\ref{sec:contamination}.  

Table~\ref{tab:AGB_counts} gives the final number of AGB candidates, the C/M ratio for the raw M32 stellar counts and our corrected values accounting for contamination from M31's population using these two approaches. 
After applying these corrections the number of carbon stars in our M32 FOV is zero or negative; indicating that there are no carbon stars in M32. This is consistent with the distribution of C-stars shown in Figure~\ref{fig:M32_Cstar_loc} which typically don't follow the general stellar distribution of M32, suggesting they are mostly from M31. A similar effect was observed by  \cite{Rowe2005} using narrow-band optical photometry, who find a lack of carbon stars in the solar-metallicity inner regions of M33.

Once contamination from M31 has been accounted for, the corrected C/M ratios (in both cases) for M32 are extremely low (consistent with zero), which is indicative of a younger, metal-rich population. 
Indeed, the M32 C/M value obtained using the Field 20 correction is lower than the C/M values for the innermost fields in M31, which have a super-solar metallicity. 
Even a 3$\sigma$ error on these values corresponds to a [Fe/H] $\sim$ $-$0.05.  
This is unexpected as the outer regions of M32 are through to have a slightly sub-solar metallicity between [Fe/H] $\sim$ $-$0.25 to $-$0.2 dex \citep{McConnachie2012, Monachesi2011, Monachesi2012} -- although the metallicity distribution inferred from optical {\em HST} data implies that there are more metal-rich stars than metal-poor ones.
However, it is also plausible that the M32 AGB population is older and metal-poor with initial masses below the limit for C-star formation.  

Integrated light estimates of M32 suggest that 20\% of the AGB light comes from C stars \citep{Davidge1990}, however the C$_2$ band at 1.77 $\mu$m was not detected. For the resolved stellar populations of M32, \citet{Jones2021} noted a lack of objects with moderately red colours in the {\em Spitzer} CMDs, where we would expect carbon stars to reside, and speculated that the intermediate-age population of M32 is older than previously thought. Although {\sc parsec-colibri} stellar isochrones from \citet{Marigo2017} and \citet{Pastorelli2019} with [M/H] $\sim$ $-$0.11 and ages between 0.6--3 Gy, corresponding to masses between 1.5--3 \Msun\ appear to be representative of these data. However, lower-mass AGB stars are below the {\em Spitzer} detection limit, which did not reach the TRGB, and there are considerable uncertainties in the estimates of this mass range. 
Finally, of the five carbon stars spectroscopically identified in the vicinity of M32, of which two are in our FOV  (SPLASH ID: 240022 and 183365, with velocities of $-$226.0 and $-$15.5 kms$^{-1}$, respectively) only one has a velocity consistent with the compact elliptical ($-275 \leq v \leq -125$ kms$^{-1}$; \citealt{Howley2013}).  Although its kinematics do not entirely exclude membership of M31. This star (SPLASH ID: 240022) is both cool and metal-rich \citep{Hamren2016}. There is no published velocity information for the spectroscopically confirmed M-stars. 


\subsection{Characterising the Age of the M32 Stellar Population}
\label{sec:Age}

\begin{figure}
\centering
\includegraphics[trim=0cm 0cm 0cm 0cm, clip=true,width=\columnwidth]{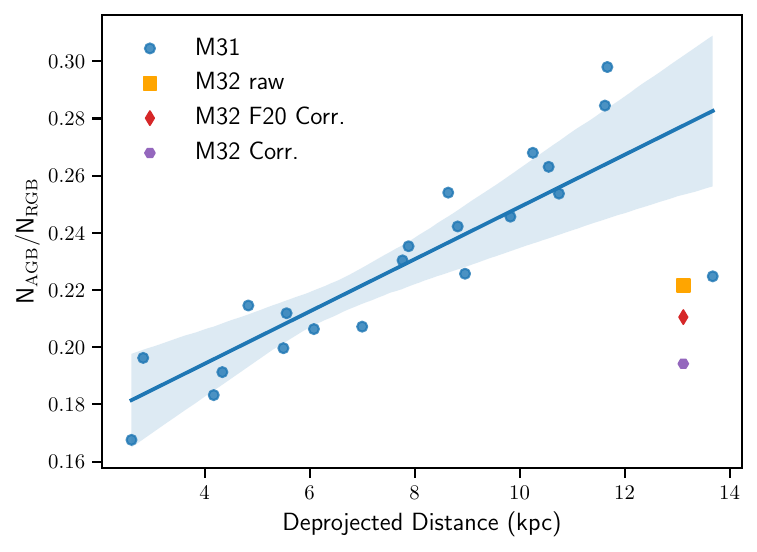}
\includegraphics[trim=0cm 0cm 0cm 0cm, clip=true,width=\columnwidth]{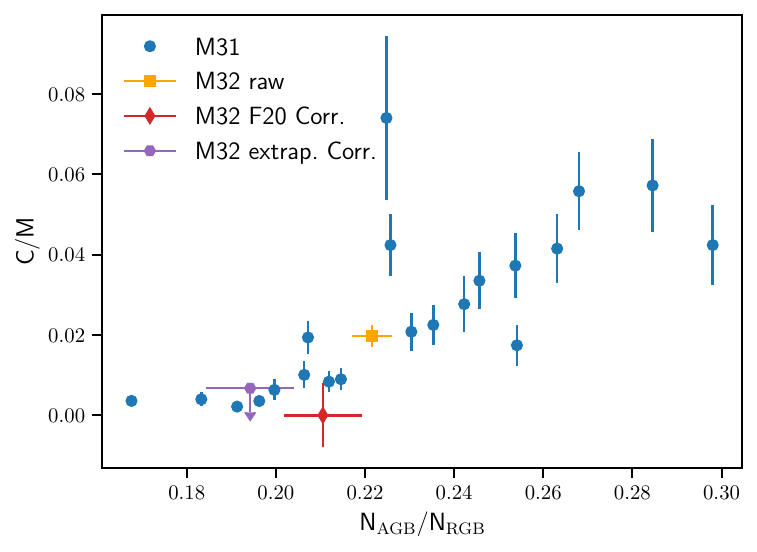}
 \caption{The ${N}_{\mathrm{AGB}}/{N}_{\mathrm{RGB}}$ can be used to infer the age of the population. Top: ${N}_{\mathrm{AGB}}/{N}_{\mathrm{RGB}}$ for M32 (orange) and the M31 (blue) fields from \citet{Boyer2019} plotted against deprojected distance from the centre of M31. 
 The effects of M31's star-forming rings on the Andromeda population can be seen at 5 and 10 kpc. The outermost M31 point is Field 20.
 Bottom:  The C/M vs.~${N}_{\mathrm{AGB}}/{N}_{\mathrm{RGB}}$ for M32 and the \citet{Boyer2019} fields in M31. The range in M32 values takes into account different correction methods to the star count numbers, with the error bars derived from Poisson statistics. The M32 values corrected using  AGB and RGB star counts of Field 20 from M31 are shown in red, while those corrected using M31 star counts extrapolated to M32's deprojected distance are in purple.} 
 \label{fig:M32_age}
 \end{figure}

The number of carbon stars present in a system is also dependent on the age of the population. Carbon stars are a short-lived phase of stellar evolution and at solar metallicity are expected to form in populations between  0.2 -- 4 Gyr \citep{Straniero1997, Marigo2007, Marigo2020}.
Outside this age range stars with lower luminosities either have not had enough time to evolve to the AGB or are older stars where the third dredge-up is not expected to occur or is inefficient due to their low core mass \citep{Sackmann1980, Karakas2002,Stancliffe2005}. For more massive stars, hot-bottom burning inhibits the formation of carbon stars. Thus the C/M values may only trace the metallicity of the system if the AGB stars populate an appropriate range in ages. 

It is possible that M32's AGB population may be older than this, with star-formation in M32 ending $\sim$2 Gyr ago \citep{Monachesi2012}. An extended and bright AGB above the TRGB strongly indicates that populations between 1 and 7 Gyr are present in our field, however, {\em Spitzer} observations of extreme AGB stars \citep{Davies2013, Jones2015a, Jones2021} are suggestive of a $\sim$3 Gyr old population. Although this is model-dependent and stellar sequences in the warm {\em Spitzer} CMDs are difficult to isolate. 
Furthermore, due to dynamical relaxation or as a result of outside-in formation the mass function may be skewed towards lower values in the outskirts of M32 compared to the nucleus \citep[e.g.,][]{HamedaniGolshan2017}.

To obtain tighter constraints on the age of the M32 population we compare the number ratio of AGB to RGB stars in our field  (with both stellar types corrected for M31 contamination) as a diagnostic for the age. 
In older populations the more massive stars have evolved off the AGB, thus higher values for ${N}_{\mathrm{AGB}}/{N}_{\mathrm{RGB}}$ indicate a younger population. 
To account for blending, photometric errors and the uncertainty on the exact location of the TRGB an interval of 0.2 to 1.2 magnitudes below the TRGB in the F127M filter and  F127M -- F153M $ >0.17$ mag defines our RGB sample; this cut was applied to both the M32 data and all the M31 fields from \citet{Boyer2019} to provide a relative age comparison. This magnitude limit was selected to ensure a high completeness factor and to minimise any possible contamination from stars at other evolutionary phases. This ratio does not take into account completeness corrections as crowding is negligible at these magnitudes and radii. 
The AGB numbers are the combined number of carbon and oxygen-rich AGB stars identified previously. The number ratio of TP-AGB and RGB stars together with their Poisson uncertainties are plotted in Figure~\ref{fig:M32_age}. 

The ${N}_{\mathrm{AGB}}/{N}_{\mathrm{RGB}}$ ratio for M32 is comparable to values from the inner disc of M31, which indicates a similar average age for both regions. 
The inner disc of M31 contains populations of stars that formed $\sim$2--4 Gyr ago \citep{Bernard2015}, it is therefore expected that M32 should also harbour stars with a similar age profile and thus have a sizeable carbon star population given its slightly sub-solar metallicity.  
On the other hand, the optical {\em HST} data of M32's RGB stars span a wide spread in colour indicating an intrinsic spread in metallicity from roughly solar to below $-$1 dex \citep{Grillmair1997, Monachesi2011}. 
In this regime disentangling the age and metallicity is difficult. 
Thus, the relatively low number of carbon stars in our field may indicate that the bright intermediate age population of M32 may have a higher metallicity than the dominant ($\sim$55\%) population of stars in the M32 elliptical galaxy which have a mean age of 5--10 Gyr.

\subsection{Dusty variable M32 stars}
\label{sec:compSpitzer}

Stars in the later stages of their AGB evolution can produce significant amounts of dust, which is expelled into the ISM via a slow dense wind. In M32, \citet{Jones2015a} identified 110 AGB stars in the mid-IR which are undergoing the ‘superwind’ phase of their evolution. During this short-lived phase of intense mass loss, the mass-loss rate exceeds the nuclear-burning rate and the remaining time a star spends on the AGB is governed by its stellar wind  \citep{VanLoon1999,Winters2000, Lagadec2008}. These `extreme’ AGB stars (x-AGB) typically dominate the dust production from low- and intermediate-mass sources in a galaxy and can account for up to 66\% of the global dust production \citep{Matsuura2009, Riebel2012, Srinivasan2009, Srinivasan2016}. In the Magellanic Clouds the vast majority of x-AGB stars produce carbon-rich dust \citep{Groenewegen2007, vanLoon2008, Ruffle2015} with $<$10\% associated with oxygen-rich dust production \citep{Jones2014, Jones2017b}. In M31, \cite{Goldman2022} found that the dust input form evolved stars in this metal-rich spiral-galaxy is dominated by oxygen-rich AGB stars. Thus, as M32 has a higher mean metallicity than the Magellanic Clouds we may expect the x-AGB dust production in the M32 elliptical to have a significant oxygen-rich component due to the higher natal abundance of oxygen and the availability of seed nuclei like SiO, MgO, Fe and TiO for efficient dust condensation \citep[e.g.,][]{Jeong2003, GailSed1999}.  

In order to establish the dust chemistry of the extreme dust producers identified in {\em Spitzer} IRAC data of M32, we match our {\em HST} catalogue to the {\em Spitzer} catalogue using a $1^{\prime\prime}$ radius.  Given the high stellar density of our field and the differences in both the depth and angular resolution of our data we restrict this matching to luminous stars that are no more than 1.5 mag below the TRGB. In the case of multiple matches to the {\em Spitzer} data, sources classified as an AGB star in the {\em HST} data were preferentially selected over the nearest positional match, as this is the more probable counterpart. This ensures reliable matches between the point-source catalogues.

Our {\em HST} data covered 62 {\em Spitzer} variables and 69 of the candidate x-AGB stars from \cite{Jones2015a, Jones2021}. 
Of the long-period variable stars matched to our {\em HST} data, 39 are M-type AGB stars and two are carbon stars in the WFC3 filters. For the x-AGB stars 45 are oxygen-rich, three are carbon stars and three have not been assigned a chemical type,  this is likely due to circumstellar dust veiling the molecular features in the photosphere.  
The remaining AGB stars identified by {\em Spitzer} as either an x-AGB or long-period variable star were either not detected by {\em HST} or are too obscured and faint to meet our {\em HST} matching criteria due to circumstellar dust extinction.

The chemically-classified x-AGB sample is predominately composed of M-type stars ($>$93\%). This is contrary to that found in metal-poor galaxies where only a small fraction of x-AGB stars are oxygen-rich \citep[e.g.,][]{Riebel2012, Srinivasan2016, Boyer2015b, Boyer2019}, and agrees with values determined for metal-rich populations \citep[e.g.,][]{Goldman2022}. Assuming the same M31 contamination ratios ($\sim$25.8\% for M-stars) for the x-AGB sample as for the general AGB population detected towards M32, our classification implies that dust-production in the outskirts of M32 is dominated by M-type stars. Furthermore, given the low number of carbon stars in our field, and the high probability that most (if not all) of these objects are associated with the disc of M31, it is likely that all AGB stars that are genuine M32 members are producing oxygen-rich dust like metal oxides and silicates, and any contribution from carbonaceous grains is negligible.

This has implications for the dust budget of M32, as previously global dust-production rates were determined using empirical relations based on the IRAC colours of x-AGB stars which were derived from sources in the Magellanic Clouds assuming a carbonaceous chemistry. Silicate grains have a lower specific opacity compared to amorphous carbon grains, thus previous estimates for the dust yields underestimate the mass return to the ISM.  
To account for this difference we derive a cumulative dust input for stars in M32 using the [3.6]--[4.5] colour–dust-production relation computed by \citet{Goldman2022} for oxygen-rich LMC AGB stars from the sample of \citet{Groenewegen2018}. Assuming all the x-AGB stars (after correcting for M31 contamination including the x-AGB stars identified as carbon-rich by our {\em HST} data) are oxygen-rich, we determine a lower limit to the cumulative dust-injection rate in M32 of 2.56 $\times 10^{-6}$ \Msuny; this increases to 1.20 $\times 10^{-5}$ \Msuny\ if we expand this to the full \cite{Jones2015a} {\em Spitzer} sample brighter than the TRGB which have [3.6]--[4.5] $>$ 0.1 mag and an 8 $\mu$m detection. This is significantly higher (by $\sim$2 orders of magnitude) than previous lower limits to the dust-production rate in M32. 
For comparison, \citet{Goldman2022} determine a cumulative dust-production rate of 2.13 $\times 10^{-4}$ \Msuny\ for oxygen-rich dust production in the PHAT M31 footprint, which covers approximately one-third of M31’s disc.

As in the Magellanic Clouds the majority of the current dust input into the ISM comes from a small number of AGB stars \citep[e.g.][]{Riebel2012, Boyer2012, Srinivasan2016}. In this instance, six sources are responsible for producing over 70\% of the dust. Using our statistical corrections it is impossible to verify their membership of M32 rather than M31, however as the resolved {\em Spitzer} photometric data excluded stars the central and inner regions ($R < 1{\buildrel{\,\prime}\over{.}}5$) of this compact high surface brightness galaxy our AGB star counts substantially underestimate the size of this evolved population and thus our lower limits are valid. 

M32 has no interstellar dust reservoirs \citep{vanDokkum1995, Gordon2006}, and whilst silicate grains have a longer residence time in the ISM compared to carbonaceous species, the dust-injection rate determined here is insufficient to replenish or produce large reservoirs of dust in this compact elliptical galaxy within a reasonable timescale. 
Especially if the intermediate-age population in M32 arises solely from a burst of star formation due to a major merger or short-lived encounter between M31 and M32. Or if the ambient radiation fields in this compact galaxy is enhanced due to its compact nature, leading to a decrease in the dust lifetime. This effect may also be seen in its gas, as M32 contains less than 4\% of the gas expected from stellar mass return \citep{Welch2001}.

\section{Conclusions}
\label{sec:conclusion}

We observed the compact elliptical galaxy M32 using medium-band WFC3/IR filters to unambiguously identify carbon- and oxygen-rich AGB stars. 
We find 57 carbon stars and 2829 M stars in our sample. 
The carbon stars appear to be almost all contaminants from M31 rather than objects intrinsic to the M32 galaxy. This is corroborated spectroscopically; the two confirmed carbon stars in our field have kinematics suggesting membership to M31 rather than M32 \citep{Hamren2016}.
We have derived C/M ratios in M32 from AGB stars selected in {\em HST} CMDs and colour--colour diagrams. The C/O ratio in M32 is $<$0.020 $\pm$ 0.003 and may reach values close to zero depending on the M31 contamination correction adopted. If carbon stars are present in M32 they are present in very low numbers. 

Fewer carbon stars may be expected in older populations. It has long been established that M32 has a strong intermediate-age (2--5 Gyr) stellar component in addition to stars older than 5 Gyr \citep{Monachesi2012} which contribute $\sim$55\% of the total mass in our fields. Comparing the ratio of AGB to RGB stars in M32 to fields across the M31 disc indicates that M32's population resembles the stellar age profile of the innermost fields from \citet{Boyer2019} that trace M31's inner disc. These fields contain stars that formed 1.5--4 Gyr ago and would be expected to produce carbon stars. 
However, if our M32 population is at the older end of this age distribution then the lack of C-stars in our field may be consistent with the narrow mass range for carbon star formation predicted from stellar evolution models by \citet{Ventura2020} for solar and super-solar metallicities. 

Finally, we used the results from the chemically-classified AGB stars to revise estimates of AGB star dust-production in M32. The dusty x-AGB candidates identified with {\em Spitzer} are predominately M-type stars, and it is likely all dust injected into the ISM of M32 is composed of silicates or metal-oxides. As silicates have a lower dust opacity compared to carbonaceous grains, this substantially increases the lower limit to the cumulative dust-production rate in M32 to $>$ 1.20 $\times 10^{-5}$ \Msuny. 
The AGB stars characterised in this study will be prime targets for follow-up spectroscopy with facilities such as {\em JWST}, as only a detailed kinematic and chemical study of the M32 stellar population will reveal its distinct components, membership and mineralogy of the ejected mass.

\section*{Acknowledgements}

OCJ has received funding from an STFC Webb fellowship.
MM acknowledges that a portion of their research was carried out at the Jet Propulsion Laboratory, California Institute of Technology, under a contract with the National Aeronautics and Space Administration (80NM0018D0004).
This research is based on observations made with the NASA/ESA Hubble Space Telescope obtained from the Space Telescope Science Institute, which is operated by the Association of Universities for Research in Astronomy, Inc., under NASA contract NAS 5–26555. These observations are associated with program 15952. Support for this work was provided by NASA through grant GO-15952.
This research made use of Astropy,\footnote{http://www.astropy.org} a community-developed core Python package for Astronomy \citep{Astropy2013} and NASA's Astrophysics Data System Bibliographic Services.

\

\noindent {\it Facilities:} {\em HST} (WFC3) - Hubble Space Telescope.

\section*{Data availability}

These data are from {\em HST} program GO-15952 (PI. Jones) and can be obtained from the
Mikulski Archive for Space Telescopes (MAST). The data underlying this article are available in the article and in its online supplementary material.




\def\aj{AJ}					
\def\actaa{Acta Astron.}                        
\def\araa{ARA\&A}				
\def\apj{ApJ}					
\def\apjl{ApJL}					
\def\apjs{ApJS}					
\def\ao{Appl.~Opt.}				
\def\apss{Ap\&SS}				
\def\aap{A\&A}					
\def\aapr{A\&A~Rev.}				
\def\aaps{A\&AS}				
\def\azh{AZh}					
\def\baas{BAAS}					
\def\jrasc{JRASC}				
\def\memras{MmRAS}				
\def\mnras{MNRAS}				
\def\pra{Phys.~Rev.~A}				
\def\prb{Phys.~Rev.~B}				
\def\prc{Phys.~Rev.~C}				
\def\prd{Phys.~Rev.~D}				
\def\pre{Phys.~Rev.~E}				
\def\prl{Phys.~Rev.~Lett.}			
\def\pasp{PASP}					
\def\pasj{PASJ}					
\def\qjras{QJRAS}				
\def\skytel{S\&T}				
\def\solphys{Sol.~Phys.}			
\def\sovast{Soviet~Ast.}			
\def\ssr{Space~Sci.~Rev.}			
\def\zap{ZAp}					
\def\nat{Nature}				
\def\iaucirc{IAU~Circ.}				
\def\aplett{Astrophys.~Lett.}			
\def\apspr{Astrophys.~Space~Phys.~Res.}		
\def\bain{Bull.~Astron.~Inst.~Netherlands}	
\def\fcp{Fund.~Cosmic~Phys.}			
\def\gca{Geochim.~Cosmochim.~Acta}		
\def\grl{Geophys.~Res.~Lett.}			
\def\jcp{J.~Chem.~Phys.}			
\def\jgr{J.~Geophys.~Res.}			
\def\jqsrt{J.~Quant.~Spec.~Radiat.~Transf.}	
\def\memsai{Mem.~Soc.~Astron.~Italiana}		
\def\nphysa{Nucl.~Phys.~A}			
\def\physrep{Phys.~Rep.}			
\def\physscr{Phys.~Scr}				
\def\planss{Planet.~Space~Sci.}			
\def\procspie{Proc.~SPIE}			
\let\astap=\aap
\let\apjlett=\apjl
\let\apjsupp=\apjs
\let\applopt=\ao


\bibliographystyle{mnras}
\bibliography{M32_HST} 







\bsp	
\label{lastpage}
\end{document}